\documentclass[10pt]{article}
\usepackage{graphicx}
\usepackage{latexsym,graphicx}
\usepackage{float}
\usepackage{amsmath}
\usepackage{amsfonts}
\usepackage{amssymb}
\usepackage{epstopdf}
\DeclareGraphicsExtensions{.eps}
\usepackage[left=2cm,top=2.5cm,right=2.5cm,bottom=1.5cm]{geometry}

\catcode `\@=11
\catcode `\@=12
\begin{document}
	\begin{center}
	\large{\bf{Transit cosmological models coupled with zero-mass scalar-field with high redshift in higher derivative theory}} \\
	\vspace{5mm}
	\normalsize{Archana Dixit$^1$, Dinesh Chandra Maurya$^2$, Anirudh Pradhan$^3$}\\
	\vspace{5mm}
	\normalsize{$^{1,3}$Department of Mathematics, Institute of Applied Sciences \& Humanities, GLA University,
			Mathura-281 406, Uttar Pradesh, India} \\
	\vspace{2mm}		
	\normalsize{$^{2}$Department of Mathematics, Faculty of Engineering \& Technology, IASE (Deemed to be University), 
	Sardarshahar-331 403, Rajsthan, India \\
		\vspace{2mm}
		$^1$E-mail:archana.dixit@gla.ac.in \\
	\vspace{2mm}
	
		$^2$E-mail:dcmaurya563@gmail.com \\
		\vspace{2mm}
		$^3$E-mail:pradhan.anirudh@gmail.com}\\
\end{center}
\vspace{5mm}
\begin{abstract}
The present study deals with a flat FRW cosmological model filled with perfect fluid coupled with the zero-mass scalar field 
in the higher derivative theory of gravity. We have obtained two types of universe models, the first one is the accelerating universe 
(power-law cosmology) and the second one is the transit phase model (hyperbolic expansion-law). We have obtained various physical 
and kinematic parameters and discussed them with observationally constrained values of $H_{0}$. The transit redshift value is 
obtained $z_{t}=0.414$ where the transit model shows signature-flipping and is consistent with recent observations. In our 
models, the present values of EoS parameter $\omega_{0}$ crosses the cosmological constant value $\omega_{0}=-1$. Also, the 
present age of the universe is calculated.
\end{abstract}



\section{Introduction}
From the recent supernovae observations ( Garnavich et al. \cite{ref1,ref2}, Perlmutter et al. \cite{ref3,ref4,ref5}, Riess et al. \cite{ref6} 
and Schmidt et al. \cite{ref7}), it was predicted that the expansion rate of the universe is increasing at present. This prediction points 
towards the presence of something (a kind of repulsive force) in the Universe which is pushing everything farther apart faster than it did 
in the early universe. The agent for this repulsive force is attributed to a mysterious entity present in the universe along with the usual 
matter and radiation. This hypothetical unknown exotic physical entity is termed as dark energy (DE). To know the reason behind this 
acceleration scenario of the present universe, there were two ways, first one to modify Einstein's field equations and another to add 
a time-dependent cosmological constant $\Lambda$. In order of modifying Einstein's general theory of relativity (GR), one is obtained 
by adding $R^{2}$ factor (square of Ricci scalar curvature) in the Einstein-Hilbert action which generalizes the Einstein's GR theory of gravity. 
Before the fruitful work of Guth, Starobinsky \cite{ref8} proposed higher derivative theories of gravity that admit inflation. But the theory 
come into focus only after the work of Guth \cite{ref9}, in which, he used a temperature phase transition mechanism. The 
good features of the higher-order theories of gravity are very interesting for a long time. Also, in literature, it is a tradition to get a 
perturbation theory by adding suitable counter terms, viz to the cosmological term $(\Lambda)$ and the $C^{ijkl}C_{ijkl}$, $R^{2}$ to the 
Einstein's action is asymptotically free, well behaved, and formally re-normalizable Stelle \cite{ref10}.  \\

The quantum field theory states a new approach to inflation and finds the dynamical history of the expanding universe through GR. Even if, 
there is a general belief among researchers that the inflation is a part of cosmological evolution. Because of this, to solve the various 
problems of inflation by reserving its importance, among the researchers have a continuous contest by presenting different modifications 
(Linde \cite{ref11}; Kolb and Turner \cite{ref12}). Chimento and Jakubi \cite{ref13} have presented cosmological models with perfect fluid 
in the context of higher derivative theory of gravity. Even if, a perfect fluid source of matter may explain matter distribution in the 
universe satisfactorily, but reason behind the accelerating scenario of the universe i.e. also, well known by the dark energy (DE) problem, 
in literature, can not be explained by perfect fluid, properly. In some literature, it is claim that the dark energy problem can be pledged 
by adding a higher derivative terms in the transport equation (Muller \cite{ref14}; Hiscock and Salmonson \cite{ref15}). The theory known by 
extended irreversible thermodynamics (EIT) is a fully relativistic formulation of the theory taking the second derivative terms in the theory, 
is developed by Israel and Stewart \cite{ref16}, and Pavon et al. \cite{ref17}, Pavon \cite{ref18}. Ratra and Peebles \cite{ref19} have obtained 
a new type of cosmological solutions in the context of GR by using transport equation the theory EIT. Pavon and Zimdahl \cite{ref20} are 
claimed that a viscous fluid source could be the presence of dark matter that causes the acceleration in expansion of the universe and 
Zimdahl et al. \cite{ref21} have claimed to the accelerating expansion due to a effective negative force driven by a cosmic anti-friction 
force. A consistency of various inflation with an arbitrary potential driving inflation are investigated by Ram \cite{ref22}. A perfect 
fluid FRW cosmological model with a minimally coupled scalar field have been investigated by Ellis and Madsen \cite{ref23}. In the case 
of minimally and non-minimally coupled scalar scalar field, a cosmological solution for the scale factor are obtained with the assumption 
the kinetic and potential terms for the scalar field are proportional to each other, Barrow and Saich \cite{ref24}, Barrow and Mimoso 
\cite{ref25} and Mimoso and Wands \cite{ref26}. Casana et al. \cite{ref27} have investigated massless DKP field in a Lyra manifold. Recently, 
Singh et al. \cite{ref28} have investigated the cosmological solutions in the presence of an imperfect fluid and zero-mass scalar field in 
higher derivative theory of gravity. Recently, Maurya and Zia \cite{ref29} have investigated Brans-Dicke scalar field cosmological model 
in Lyra's Geometry. Some authors \cite{ref30}-\cite{ref33} have studied and discussed zero-mass scalar field in different 
contexts.\\

Motivated by above discussion, we have considered a flat FRW universe filled with perfect fluid coupled with the zero-mass scalar field in 
higher derivative theory of gravity and obtained a transit phase model. The out lines of the paper as follows: Section $1$ is introductory, 
Section $2$ introduces the metric and field equations. Solutions and discussion of results are given in Section $3$ and finally conclusions 
are given in Section $4$. 

\section{Metric and Field Equations}
To investigate the physical behaviour of the universe, we consider the action principle as
\begin{equation}\label{1}
I=\int{\left[\frac{1}{2}f(R)+L_{m}\right]}\sqrt{-g}dx^{4}
\end{equation}
where $f(R)$ is a function of Ricci scalar curvature $R$ and its higher powers, $g$ is the determinant of the metric 
tensor $g_{ij}$. $L_{m}$ is matter Lagrangian, in cosmic unit taking $8\pi G=c=1$. By the variation of action $(1)$ with respect 
to metric field $g_{ij}$ and scalar field $\phi$ respectively, we get
\begin{equation}\label{2}
f'(R)R_{ij}-\frac{1}{2}f(R)g_{ij}+f''(R)(\nabla_{i}\nabla_{j}R-\Box Rg_{ij})+f'''(R)(\nabla_{i}R\nabla_{j}R-\nabla^{k}R\nabla_{k}Rg_{ij})=-T_{ij}
\end{equation}
and
\begin{equation}\label{3}
\sqrt{-g}\phi_{,i}^{,i}=(\sqrt{-g}g^{ij}\phi_{,i})_{,j}=0
\end{equation}
where $\Box=g_{ij}\nabla^{i}\nabla^{j}$ and $\nabla_{i}$ is the covariant differential operator, and prime represents the derivative 
with respect to $R$. The Stress-energy momentum tensor for the fluid coupled with zero-mass scalar-field is given as
\begin{equation}\label{4}
T_{ij}=T^{F}_{ij}+T^{\phi}_{ij}
\end{equation}
where
\begin{equation}\label{5}
T^{F}_{ij}=(\rho+p)u_{i}u_{j}+pg_{ij}
\end{equation}
and
\begin{equation}\label{6}
T^{\phi}_{ij}=\phi_{,i}\phi_{,j}-\frac{1}{2}g_{ij}\phi_{,k}\phi^{,k}
\end{equation}
where $\rho$ is the energy density and $p$ is the isotropic pressure of the fluid. In a co-moving coordinate system, the four-velocity 
vector $u_{i}$ are satisfies the condition $u_{i}u^{i}=-1$. $\phi$ is the zero-mass scalar-field.\\

Now, we have considered the flat Robertson-Walker spacetime metric, given by
\begin{equation}\label{7}
ds^{2}=-dt^{2}+a^{2}(t)[dr^{2}+r^{2}(d\theta^{2}+sin^{2}\theta d\varphi^{2})]
\end{equation}
where $a(t)$ is the scale factor of the universe. The scalar curvature $R$ and trace $T$ of the energy momentum tensor $T_{ij}$ of the 
universe are given as respectively
\begin{equation}\label{8}
R=-6(\dot{H}+2H^{2})
\end{equation}
and
\begin{equation}\label{9}
T=-(\rho-3p)-\dot{\phi}^{2}
\end{equation}
where $H=\frac{\dot{a}}{a}$ is the Hubble parameter and an over dot represents a derivative with respect to time. The trace and $(0,0)$ 
components of $(2)$ are respectively, given by
\begin{equation}\label{10}
f'(R)R-2f(R)-3f''(R)(\ddot{R}+3H\dot{R})-3f'''(R)\dot{R}^{2}+T=0
\end{equation}
and
\begin{equation}\label{11}
f'(R)R^{0}_{0}-\frac{1}{2}f(R)-3f''(R)H\dot{R}+T^{0}_{0}=0
\end{equation}
Now, since, we have considered a higher derivative theory of gravity defined by $f(R)=R+\alpha R^{2}$. Using Eq. $(8)$ in $(10)$ and $(11)$, we obtain
\begin{equation}\label{12}
\dot{H}+2H^{2}+6\alpha [\dddot{H}+7H\ddot{H}+4\dot{H}^{2}+12H^{2}\dot{H}]=-\frac{1}{6}T
\end{equation}
and
\begin{equation}\label{13}
3H^{2}+18\alpha[2H\ddot{H}+\dot{H}^{2}+10H^{2}\dot{H}]=\rho+\frac{\dot{\phi}^{2}}{2}
\end{equation}
From Eq. $(3)$, we get
\begin{equation}\label{14}
\ddot{\phi}+3H\dot{\phi}=0
\end{equation}
Taking covariant derivative of the field equation $(2)$ and applying energy conservation laws, we obtain
\begin{equation}\label{15}
36\alpha(\ddot{H}+4H\dot{H})(2\dot{H}+H^{2})=0
\end{equation}
and
\begin{equation}\label{16}
\dot{\rho}+3H(\rho+p)=0
\end{equation}
where over dot denotes ordinary derivatives with respect to cosmic time $`t'$.
\section{Solutions of the field equations and discussion of results}
From redshift relation $\frac{a(t_{0})}{a(t)}=1+z$ with $a(t_{0})=1$, we have found
\begin{equation}\label{17}
\dot{H}=-(1+z)HH' \hspace{15.5cm}
\end{equation}
\begin{equation}\label{18}
\ddot{H}=(1+z)H^{2}H'+(1+z)HH'^{2}+(1+z)^{2}H^{2}H''   \hspace{10cm}
\end{equation}
\begin{equation}\label{19}
\dddot{H}=-(1+z)H^{3}H'-4(1+z)^{2}H^{2}H'^{2}-3(1+z)^{2}H^{3}H''-(1+z)^{3}H^{2}H'^{3}-4(1+z)^{3}H^{2}H'H''-(1+z)^{3}H^{3}H'''
\end{equation}
where $H'=\frac{dH}{dz}$.\\
Now, integrating Eq. $(14)$, we have obtained the expression for zero-mass scalar field as
\begin{equation}\label{20}
\dot{\phi}=\frac{n}{a^{3}}=n(1+z)^{3}
\end{equation}
or
\begin{equation}\label{21}
\phi=-n\int\frac{(1+z)^{2}}{H}dz
\end{equation}
where $n$ is an integrating constant.\\
Therefore, using these results in Eqs. $(12)$ and $(13)$ we obtain 
\begin{equation}\label{22}
T=6(1+z)HH'-12H^{2}+36\alpha(1+z)H^{2}[6HH'-4(1+z)HH''+(1+z)^{2}H'^{3}+4(1+z)^{2}H'H''+(1+z)^{2}HH'''-7H'^{2}]
\end{equation}
and
\begin{equation}\label{23}
\rho=3H^{2}+18\alpha(1+z)H^{2}[(3+z)H'^{2}+2(1+z)HH''-8HH']-\frac{n^{2}}{2}(1+z)^{6}
\end{equation}
If the equation of state (EoS) for the considered fluid is read as $p=\rho \omega$, then EoS parameter $\omega$ is calculated from Eq. $(16)$ as
\begin{equation}\label{24}
\omega=-1-\frac{\dot{\rho}}{3H\rho}
\end{equation}
or
\[
2(1+z)HH'-n^{2}(1+z)^{6}
\]
\begin{equation}\label{25}
\omega=-1+\frac{-12\alpha(1+z)H[2(1+z)(2HH'^{2}-H'^{3}+H^{2}H'')-(1+z)^{2}H(H'H''+H'^{3}+HH''')+4H^{2}H'+5HH'^{2}]}{[3H^{2}+
18\alpha(1+z)H^{2}[(3+z)H'^{2}+2(1+z)HH''-8HH']-\frac{n^{2}}{2}(1+z)^{6}]}
\end{equation}
Now, from Eq. $(15)$, we have
\begin{equation}\label{26}
\ddot{H}+4H\dot{H}=0 ~~~~~\text{or}~~~~2\dot{H}+H^{2}=0
\end{equation}
\begin{equation}\label{27}
2\dot{H}+H^{2}=0~\implies~H=\frac{k}{\sqrt{a}}=k(1+z)^{\frac{1}{2}}
\end{equation}
and
\begin{equation}\label{28}
\ddot{H}+4H\dot{H}=0~\implies ~\dot{H}=\frac{c_{1}}{a^{4}}=c_{1}(1+z)^{4}
\end{equation}
where $k$ and $c_{1}$ are arbitrary integrating constant.\\
Thus, there are two cases to obtain the Hubble parameter and scale-factor which gives two different types of universe model.
\subsection{Accelerating Model: Case-I: when $H=\frac{k}{\sqrt{a}}$}
\begin{equation}\label{29}
H=\frac{k}{\sqrt{a}}~\implies~a(t)=\left(\frac{kt+l}{2}\right)^{2} 
\end{equation}
which shows power-law cosmology and also, in terms of redshift, the Hubble parameter $(H)$ obtained as
\begin{equation}\label{30}
H=k(1+z)^{\frac{1}{2}}
\end{equation}
It is known that at present $z=0$ putting in Eq. $(30)$, we get the present value of Hubble parameter $H_{0}=k$. Recently, Maurya 
and Zia \cite{ref29} estimated the present value of the Hubble parameter $H_{0}=71.27$. Figure $1$ represents the variation of Hubble 
parameter $H(z)$ over redshift $z$. We see that $H(z)$ is an increasing function of redshift $z$ which reveals the expanding nature of 
the universe.\\

The deceleration parameter for the model is obtained as a constant negative value $q=-0.5$ that shows that the accelerating 
scenario of the universe.\\

Now, Eq. $(21)$ becomes
\begin{equation}\label{31}
\phi=\phi_{0}+\frac{2n}{5k}[1-(1+z)^{\frac{5}{2}}]
\end{equation}
where $\phi_{0}$ is the present value of scalar field $\phi$. Figures 2(a) and 2(b) represent the variation of scalar field $\phi(z)$ 
versus redshift $z$.\\

\begin{figure}[H]
	\centering
	\includegraphics[width=10cm,height=8cm,angle=0]{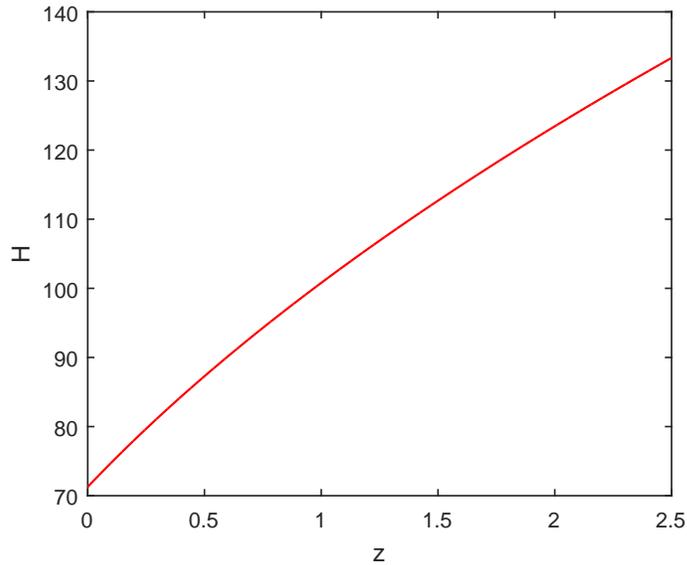}
	\caption{The plot of Hubble parameter $H(z)$ versus redshift $z$ for $H_{0}=71.27$.}
\end{figure}
\begin{figure}[H]
	(a)\includegraphics[width=8cm,height=7cm,angle=0]{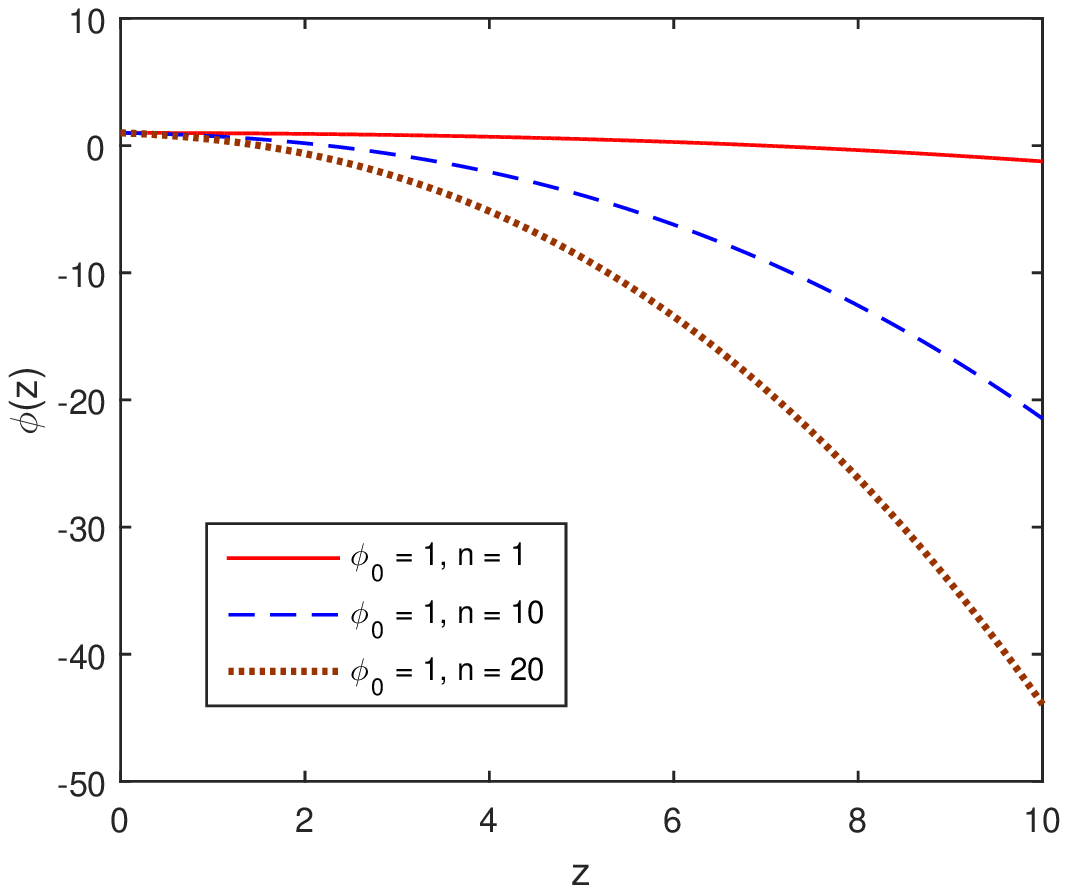}
	(b)\includegraphics[width=8cm,height=7cm,angle=0]{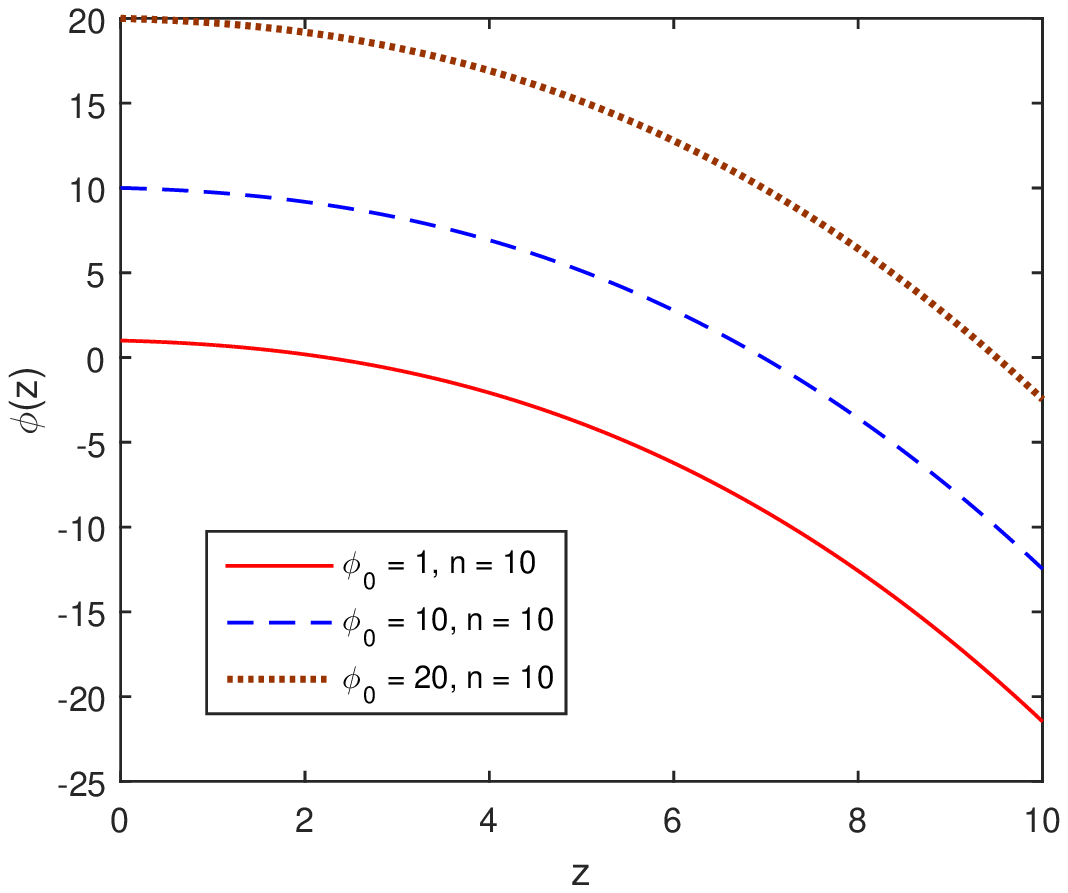}
	\caption{The plots of scalar field $\phi(z)$ versus redshift $z$ for $H_{0}=71.27$.}
\end{figure}
Using Eqs. $(30)$ and $(31)$ in Eqs. $(22)-(25)$, we obtain the expressions for the trace $T$, energy density $\rho$ of the fluid, 
and EoS parameter $\omega$ as
\begin{equation}\label{32}
T=-9k^{2}(1+z)+\frac{9\alpha k^{4}}{2}(1+z)^{2}\left[31+(1+z)^{\frac{1}{2}}-\frac{14k^{2}}{1+z}\right] 
\end{equation}
\begin{equation}\label{33}
\rho=3k^{2}(1+z)-\frac{9\alpha k^{4}}{2}(1+z)(22+23z)-\frac{n^{2}}{2}(1+z)^{6}
\end{equation}
\begin{equation}\label{34}
\omega=-1+\frac{1-\frac{n^{2}}{k^{2}}(1+z)^{5}-\frac{3}{2}\alpha k^{2}(1+z)\left[2(1+2z)+\frac{8}{1+z}-
\frac{k}{(1+z)^{\frac{1}{2}}}+16\right] }{3-\frac{9\alpha k^{2}}{2}(22+23z)-\frac{n^{2}}{2k^{2}}(1+z)^{5}}
\end{equation}
Eq. $(32)$ represents the expression for trace of the total energy momentum tensor $T_{ij}$ and Figure $3$ represents the plot of the 
trace $T$ over the variation of redshift $z$ for the different values of $\alpha$ and constrained by the Hubble constant $H_{0}$. 
At present $z=0$, $T_{0}=-9k^{2}+144\alpha k^{4}-63\alpha k^{6}$ which depends on the value of $k$ and $\alpha$.\\

Eq. $(33)$ represents the expression for energy density $\rho$ in terms of redshift $z$ and Figures 4(a), 4(b), and 4(c) show the variation 
of energy density $\rho(z)$ over the variation of redshift $z$ constrained by Hubble constant $H_{0}$ for different choices of the value 
of $\alpha$ and $n$. At present $z=0$, $\rho_{0}=3k^{2}(1-33\alpha k^{2})-\frac{n^{2}}{2}$ and in past as $z$ increases, $\rho$ increases 
to a large value. Here, $\rho_{0} > 0$ for $3k^{2}(1-33\alpha k^{2}) > \frac{n^{2}}{2}$.\\

Eq. $(34)$ denotes the expression for EoS parameter $\omega(z)$ and Figure 5(a), 5(b), and 5(c) represent the evolution of EoS parameter 
$\omega$ over the variation of redshift $z$ constrained by $H_{0}$. One can see that at present $z=0$,
\begin{equation}\nonumber
\omega_{0}=-1+\frac{2k^{2}-78\alpha k^{4}+3\alpha k^{5}-2n^{2}}{6k^{2}-198\alpha k^{4}-n^{2}}
\end{equation}
and $\lim\limits_{z\to\infty}\omega(z)=1$ which are consistent with recent observations \cite{ref1}-\cite{ref7}.
\begin{figure}[H]
	\centering
	\includegraphics[width=10cm,height=8cm,angle=0]{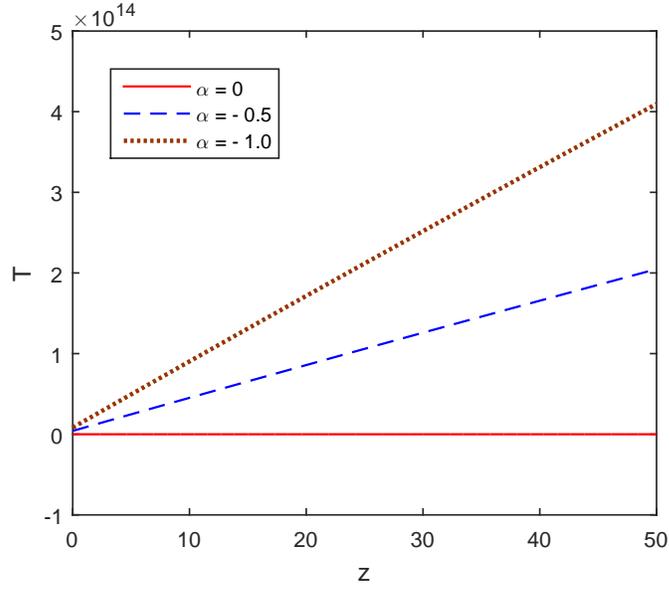}
	\caption{The plots of trace $T(z)$ versus redshift $z$ for $H_{0}=71.27$.}
\end{figure}
\begin{figure}[H]
	(a)\includegraphics[width=5cm,height=5cm,angle=0]{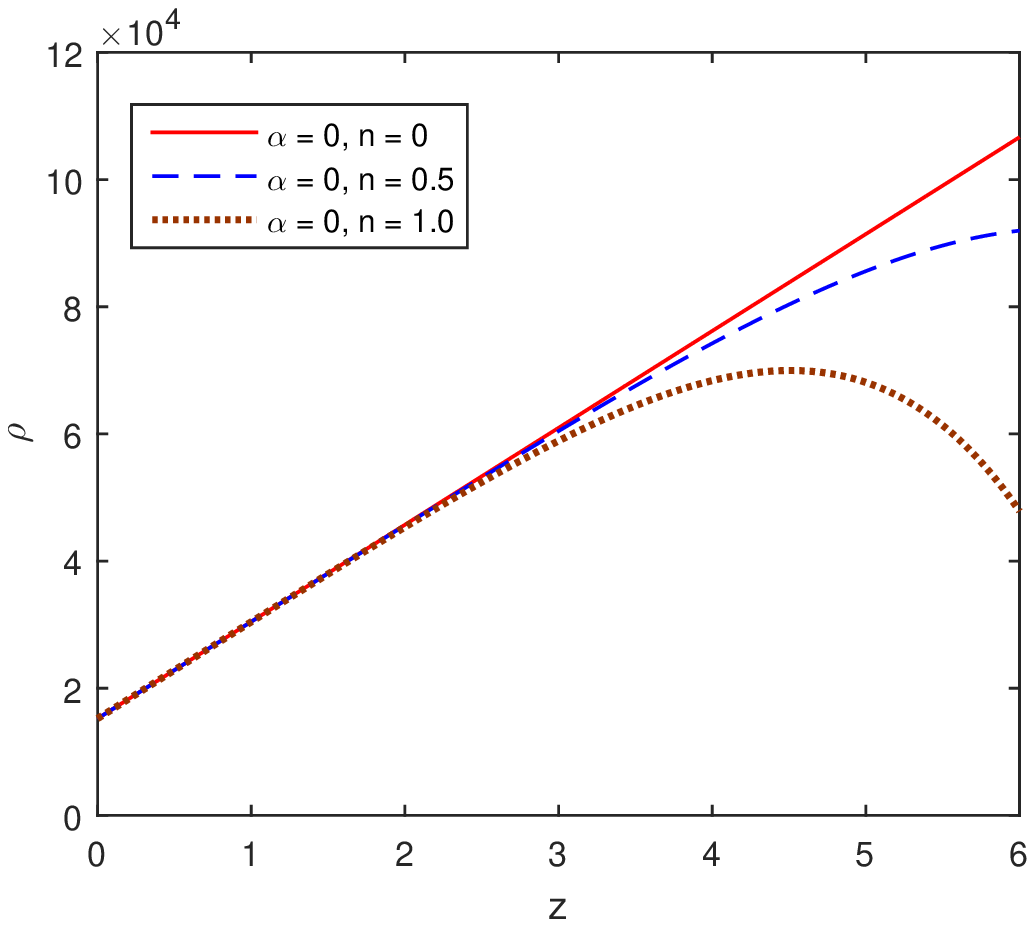}
	(b)\includegraphics[width=5cm,height=5cm,angle=0]{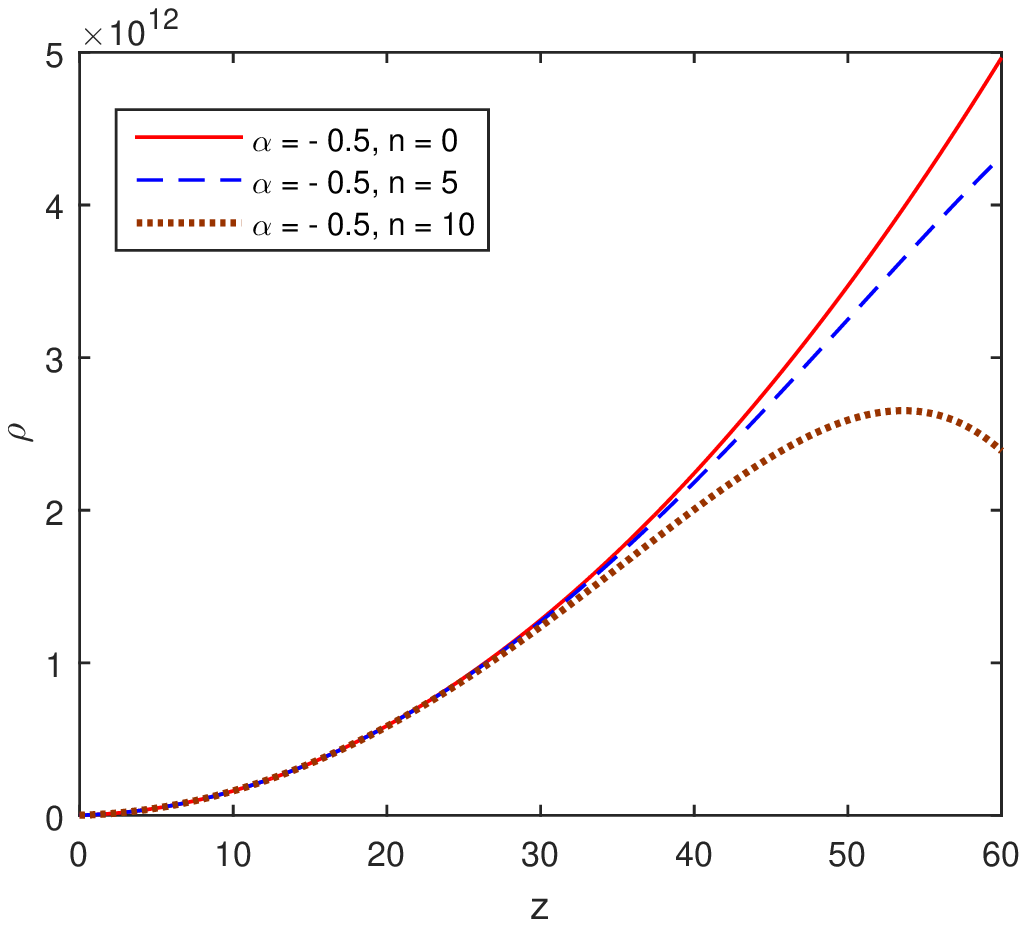}
	(c)\includegraphics[width=5cm,height=5cm,angle=0]{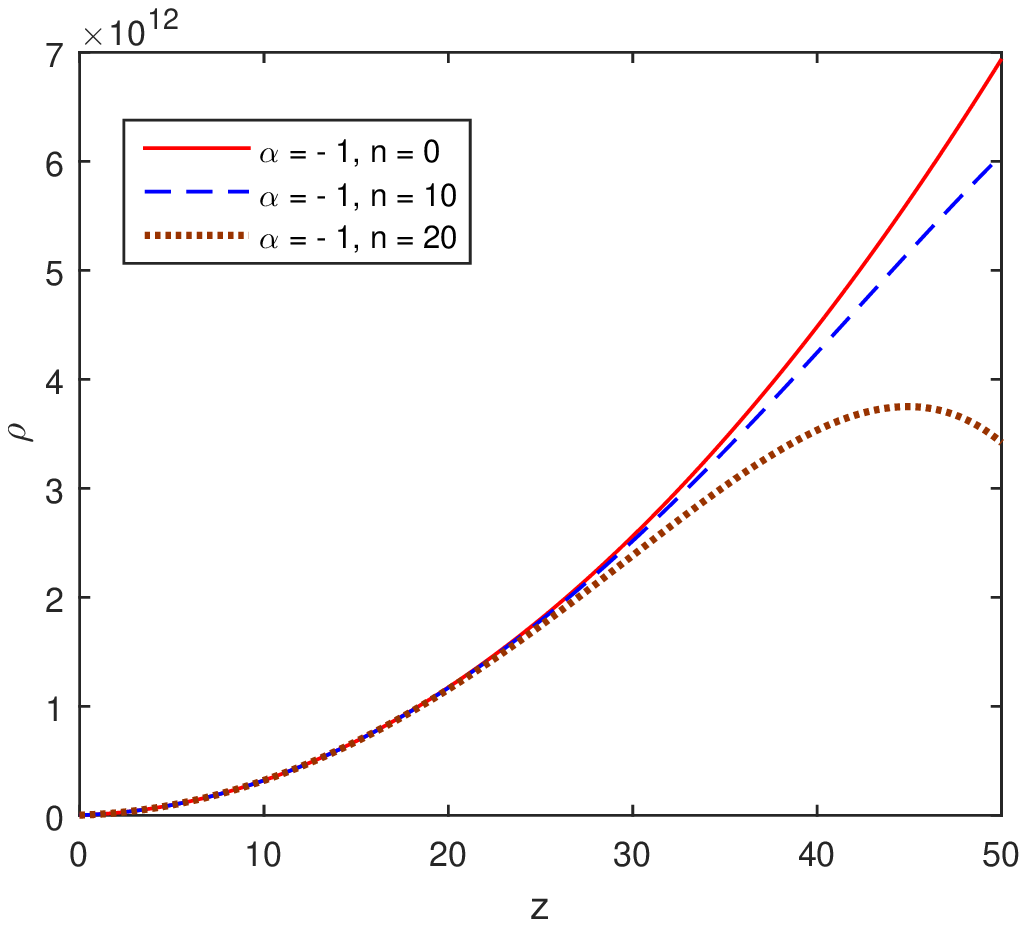}
	\caption{The plots of energy density $\rho(z)$ versus redshift $z$ for $H_{0}=71.27$.}
\end{figure}

\begin{figure}[H]
	(a)\includegraphics[width=6cm,height=5cm,angle=0]{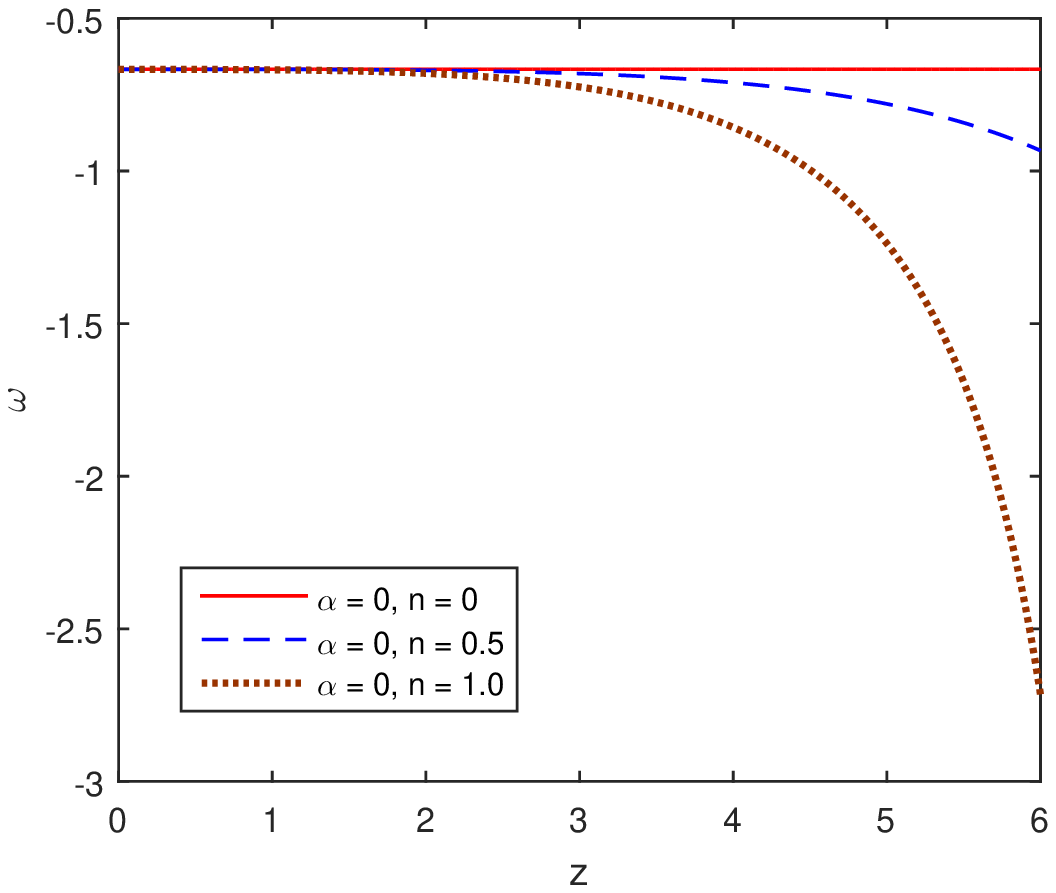}
	(b)\includegraphics[width=6cm,height=5cm,angle=0]{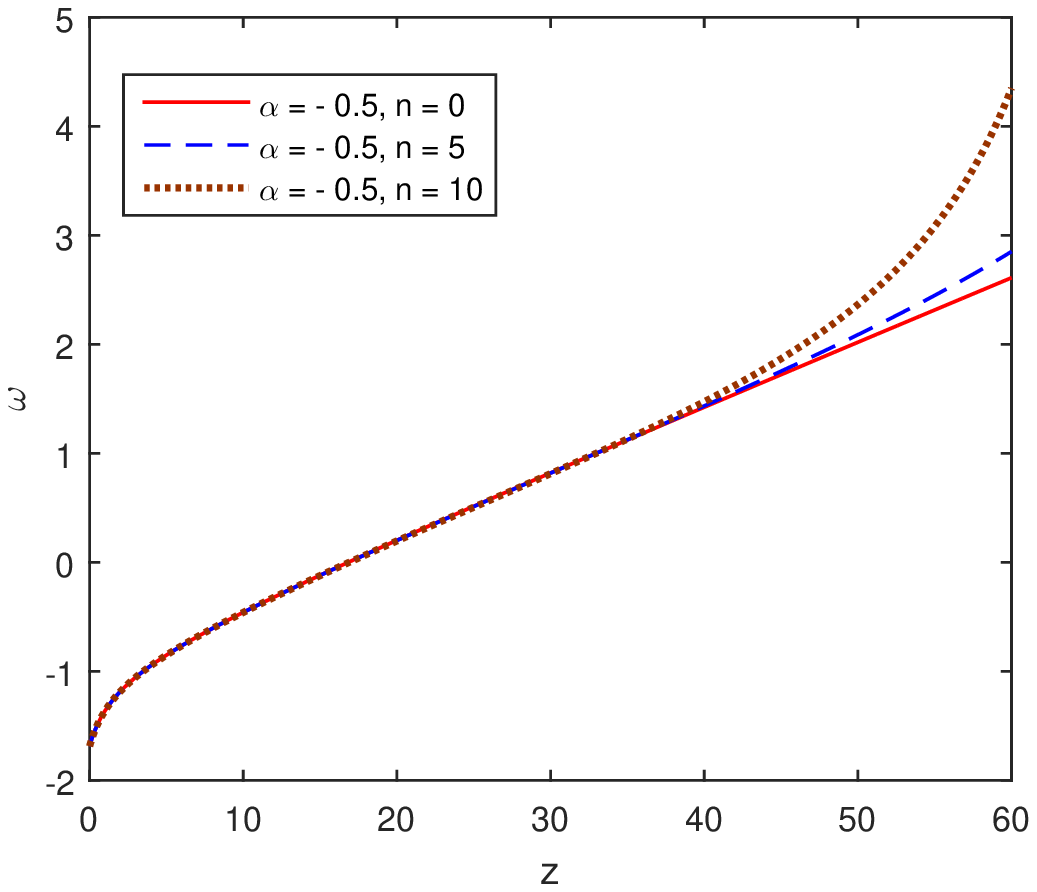}
	(c)\includegraphics[width=6cm,height=5cm,angle=0]{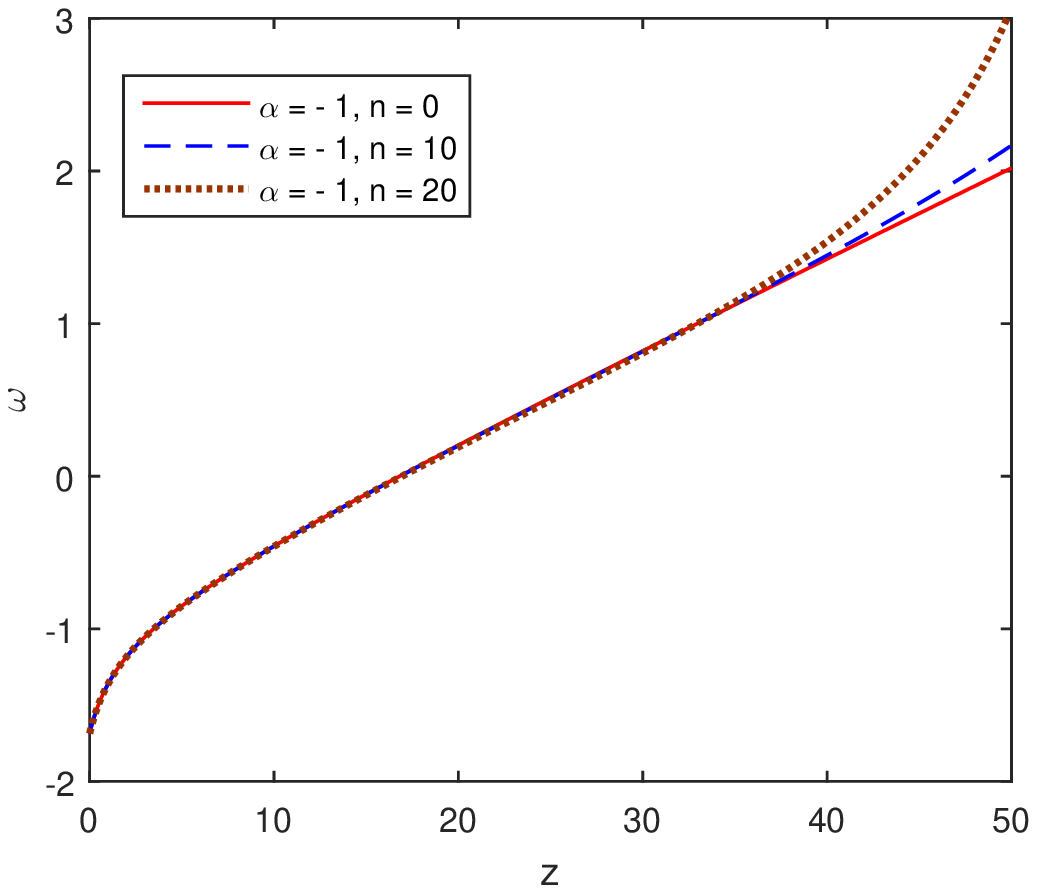}
	\caption{The plots of EoS parameter $\omega(z)$ versus redshift $z$ for $H_{0}=71.27$.}
\end{figure}
\subsection*{Age of the Universe}
The age of the cosmic universe is calculated as
\begin{equation}\label{35}
t_{0}-t=\int_{0}^{z}\frac{dz}{(1+z)H(z)}
\end{equation}
\begin{equation}\label{36}
H_{0}(t_{0}-t)=2\left[1-\frac{1}{\sqrt{1+z}}\right] 
\end{equation}
\begin{figure}[H]
	\centering
	\includegraphics[width=10cm,height=8cm,angle=0]{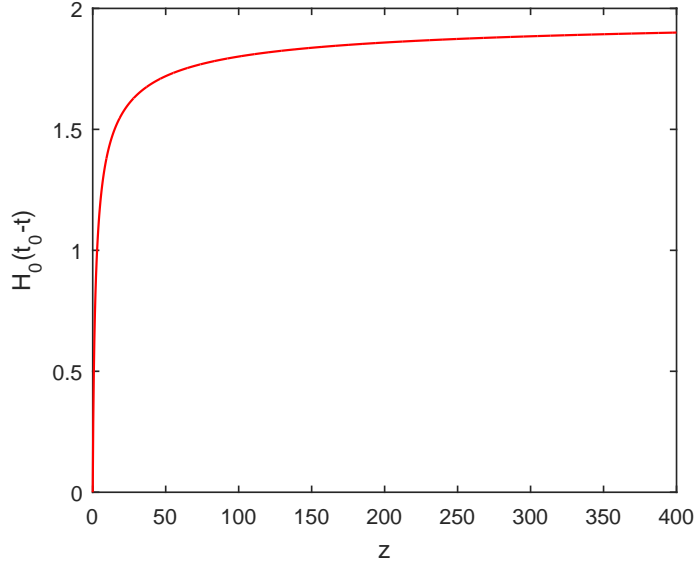}
	\caption{The plot of cosmic time $t$ versus redshift $z$}
\end{figure}
From Figure $6$, the constant graph of time shows the age of the universe i.e. $\lim\limits_{z\to\infty}H_{0}(t_{0}-t)=2\implies H_{0}t_{0}=2$ 
and hence, the present age of the universe $t_{0}=27.44 Gyrs$.
\subsection{Transit Model: Case-II: when $\dot{H}=\frac{c_{1}}{a^{4}}$}
From equations $\dot{H}=\frac{c_{1}}{a^{4}} ~~\text{and}~~\dot{H}=-(1+z)H\frac{dH}{dz}$ we have obtained the Hubble parameter $H$ as
\begin{equation}\label{37}
H=[2c_{2}-\frac{c_{1}}{2}(1+z)^{4}]^{\frac{1}{2}}
\end{equation}
where $c_{1}$ and $c_{2}$ are arbitrary constants. At present $z=0$, we found the $H_{0}=[2c_{2}-\frac{c_{1}}{2}]^{\frac{1}{2}}$. 
Figure $7$ represents its variation over redshift $z$.\\

Now, we have 
\begin{equation}\label{38}
\frac{\dot{a}}{a}=\sqrt{2c_{2}-\frac{c_{1}}{2a^{4}}}
\end{equation}
Let us choose the arbitrary constants $c_{1}, c_{2}$ as $c_{2}=-c_{1}=m_{1}$ and $c_{2}=c_{1}=m_{1}$, in these cases, integrating 
Eq. $(38)$, we get the scale factor as
\begin{equation}\label{39}
a(t)=\frac{1}{\sqrt{2}}[sinh(2\sqrt{2}m_{1}t+4m_{2})]^{\frac{1}{2}}
\end{equation}
and
\begin{equation}\label{40}
a(t)=\frac{1}{\sqrt{2}}[cosh(2\sqrt{2}m_{1}t+4m_{2})]^{\frac{1}{2}}
\end{equation}
respectively. Where $m_{1}, m_{2}$ are integrating constants. Here, we have considered first case.\\
Now, we have calculated the deceleration parameter $q(z)$ as
\begin{equation}\label{41}
q=\frac{d}{dt}\left(\frac{1}{H}\right)-1~~\implies~~q(z)=-\frac{4c_{2}+c_{1}(1+z)^{4}}{4c_{2}-c_{1}(1+z)^{4}} 
\end{equation}
At present $z=0$, $q_{0}=-\frac{4c_{2}+c_{1}}{4c_{2}-c_{1}}$ and in the past as $z\to\infty$, we have $\lim\limits_{z\to\infty}q(z)=1$. 
Also, Figure 7(b) represents the variation of deceleration parameter $q(z)$ over the variation of redshift $z$. One can see that $q(z)$ 
is an increasing function of redshift $z$ and at present $z=0$, $q_{0}=-0.6$ and at $z_{t}=0.414$ expansion transits its phase from 
deceleration to acceleration. Thus, our universe model is a transit phase model.\\
\begin{figure}[H]
	(a)\includegraphics[width=8cm,height=7cm,angle=0]{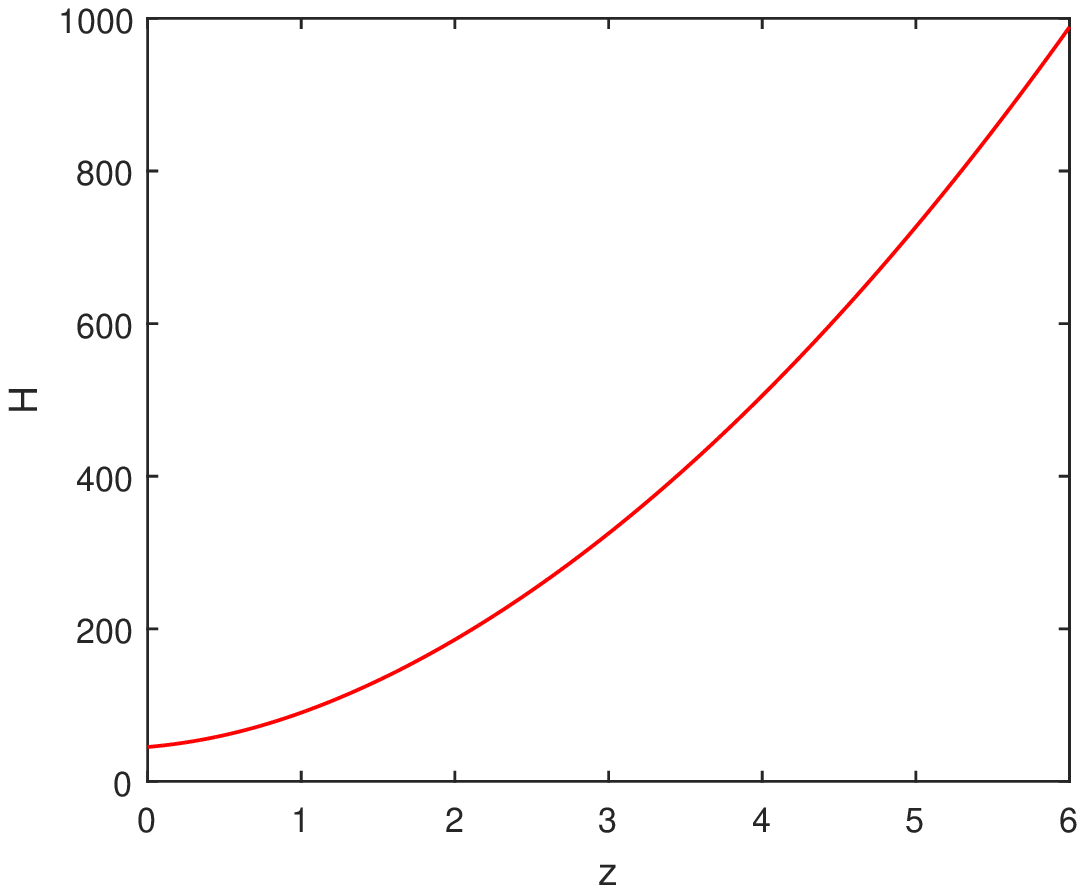}
	(b)\includegraphics[width=8cm,height=7cm,angle=0]{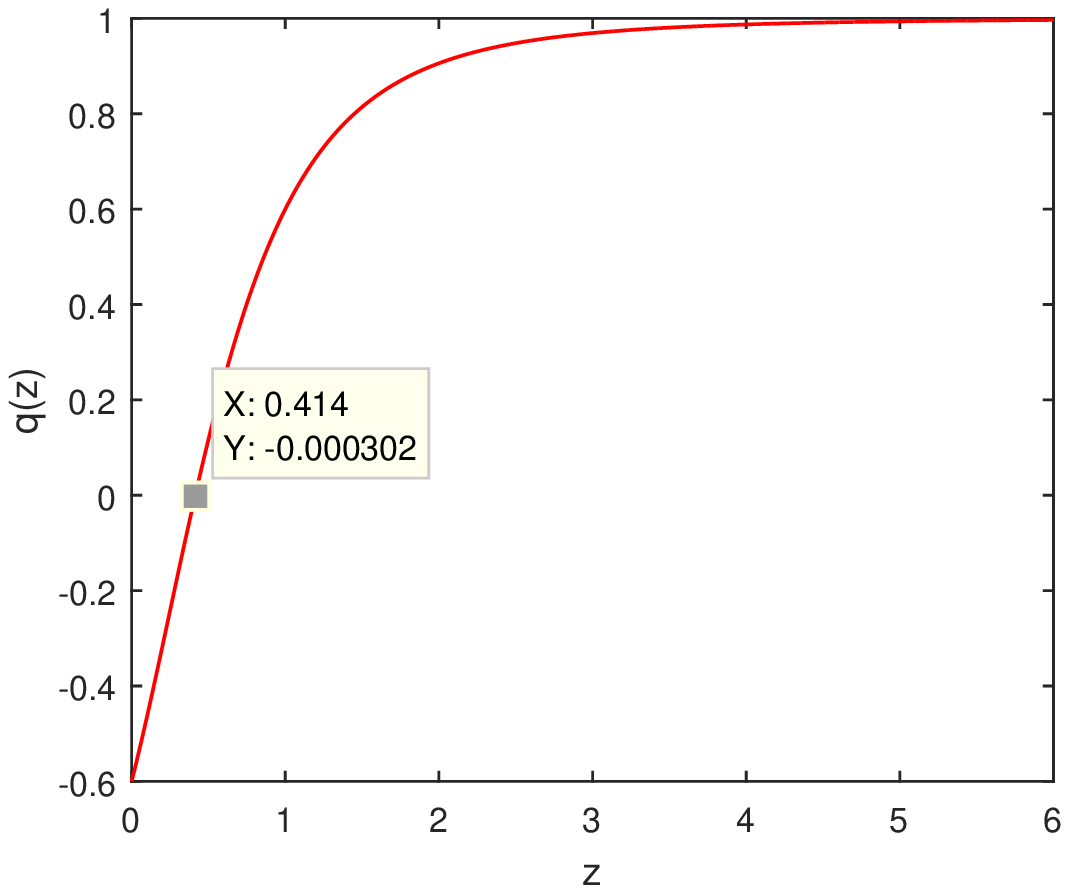}
	\caption{(a)The plot of Hubble parameter $H$ versus redshift $z$ and (b) The plot of deceleration parameter $q(z)$ versus 
	redshift $z$ for $c_{1}=-\frac{2}{5}H_{0}^{2},~c_{2}=\frac{2}{5}H_{0}^{2}$.}
\end{figure}
Now, using Eq. $(37)$ in Eqs. $(22)-(25)$, we obtain trace $T$, energy density $\rho$ and EoS parameter $\omega$ in terms of redshift $z$ as
\begin{equation}\label{42}
T=-24c_{2}-\frac{36\alpha c_{1}^{2}(1+z)^{8}}{[2c_{2}-\frac{c_{1}}{2}(1+z)^{4}]}[3-2c_{2}-3c_{1}(1+z)^{4}+c_{1}(1+z)^{4}[2c_{2}-
\frac{c_{1}}{2}(1+z)^{4}]^{\frac{1}{2}}]
\end{equation}
\begin{equation}\label{43}
\rho=6c_{2}-\frac{3c_{1}}{2}(1+z)^{4}+36\alpha c_{1}(1+z)^{4}[2c_{2}+c_{1}(1+z)^{3}-2c_{1}(1+z)^{4}]-\frac{n^{2}}{2}(1+z)^{6}
\end{equation}
\begin{equation}\label{44}
\omega=-1+\frac{L(z)}{\rho(z)}
\end{equation}
where
\[
L(z)=-2c_{1}(1+z)^{4}+\frac{12\alpha c_{1}(1+z)^{3}}{[2c_{2}-\frac{c_{1}}{2}(1+z)^{4}]}\times
\]
\begin{equation}\label{45}
[4c_{2}(5-6c_{2})(1+z)+24c_{1}c_{2}(1+z)^{5}-2c_{1}^{2}(1+z)^{8}+\frac{3c_{1}^{2}}{2}(1+z)^{9}+[30c_{2}+
\frac{5}{2}c_{1}(1+z)^{4}-c_{1}^{2}(1+z)^{9}][2c_{2}-\frac{c_{1}}{2}(1+z)^{4}]^{\frac{1}{2}}]-n^{2}(1+z)^{6}
\end{equation}
Eq. $(42)$ represents the expression for trace $T(z)$ of the total energy momentum tensor $T_{ij}$ and Figure $8$ represents the 
plot of the trace $T$ over the variation of redshift $z$ for the different values of $\alpha$ and constrained by the Hubble constant 
$H_{0}$. At present $z=0$, $T_{0}=-24c_{2}-\frac{72\alpha c_{1}^{2}}{4c_{2}-c_{1}}[3(1-c_{1})-2c_{2}+c_{1}(2c_{2}-
\frac{c_{1}}{2})^{\frac{1}{2}}]$ which depends on the value of $c_{1}, c_{2}$ and $\alpha$.\\

Eq. $(43)$ represents the expression for energy density $\rho(z)$ in terms of redshift $z$ and Figures 9(a), 9(b), and 9(c) show the 
variation of energy density $\rho(z)$ over the variation of redshift $z$ constrained by Hubble constant $H_{0}$ for different choices 
of the value of $\alpha$ and $n$. At present $z=0$, $\rho_{0}=6c_{2}-\frac{3}{2}c_{1}+36\alpha c_{1}(2c_{2}-c_{1})-\frac{n^{2}}{2}$ and 
in past as $z$ increases, $\rho$ increases to a large value.\\

Eq. $(44)$ denotes the expression for EoS parameter $\omega(z)$ and Figures 10(a), 10(b), and 10(c) represent the evolution of EoS 
parameter $\omega$ over the variation of redshift $z$ constrained by $H_{0}$. One can see that at present $z=0$,
\begin{equation}\nonumber
\omega_{0}=-1+\frac{-2c_{1}+\frac{24\alpha c_{1}}{4c_{2}-c_{1}}[4c_{2}(5-6c_{2})+24c_{1}c_{2}-\frac{1}{2}c_{1}^{2}+(30c_{2}+
\frac{5}{2}c_{1}-c_{1}^{2})(2c_{2}-\frac{1}{2}c_{1})^{\frac{1}{2}}]-n^{2}}{6c_{2}-\frac{3}{2}c_{1}+36\alpha c_{1}(2c_{2}-c_{1})-\frac{n^{2}}{2}}
\end{equation}
and $\lim\limits_{z\to\infty}\omega(z)=\infty$ which are consistent with recent observations \cite{ref1}-\cite{ref7}.
\begin{figure}[H]
	\centering
	\includegraphics[width=10cm,height=8cm,angle=0]{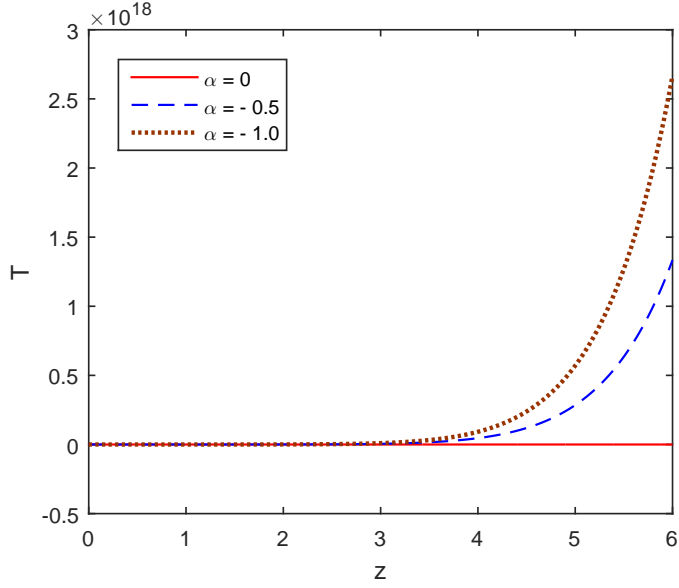}
	\caption{The plots of trace $T(z)$ versus redshift $z$ for $H_{0}=71.27$.}
\end{figure}
\begin{figure}[H]
	(a)\includegraphics[width=6cm,height=5cm,angle=0]{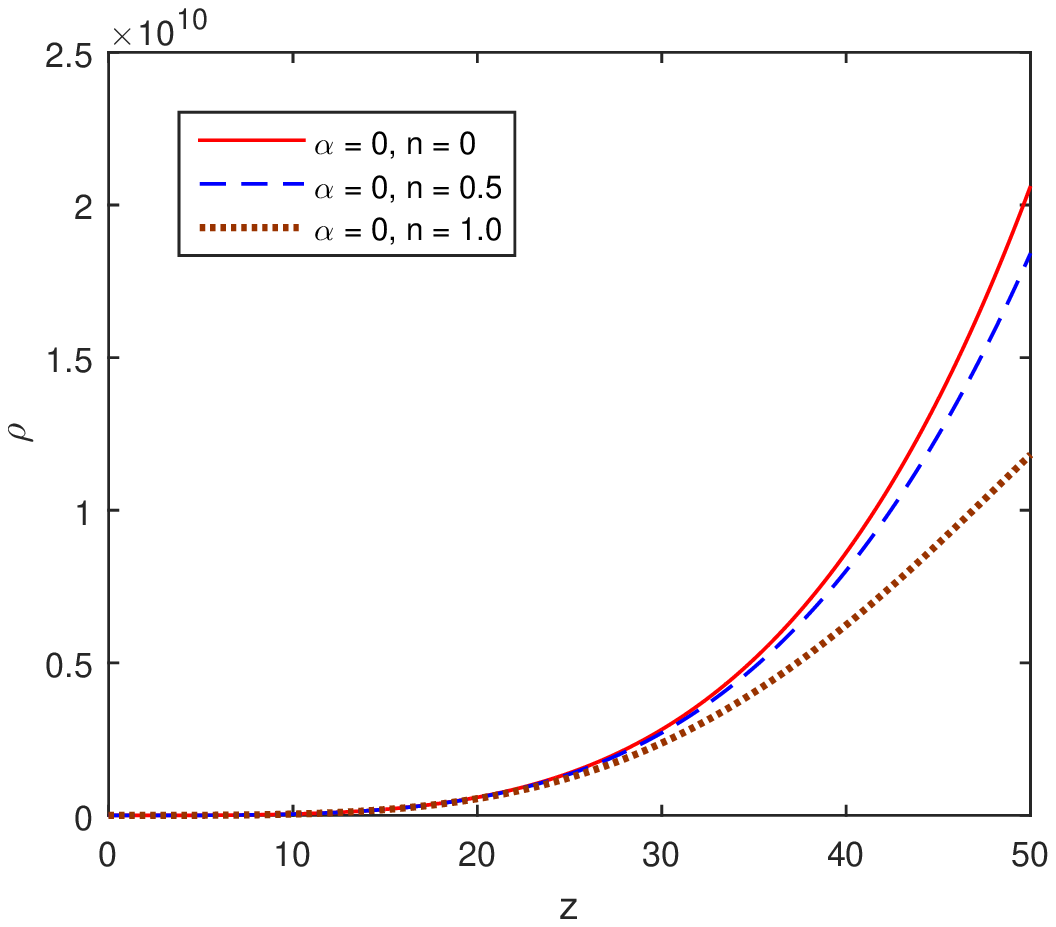}
	(b)\includegraphics[width=6cm,height=5cm,angle=0]{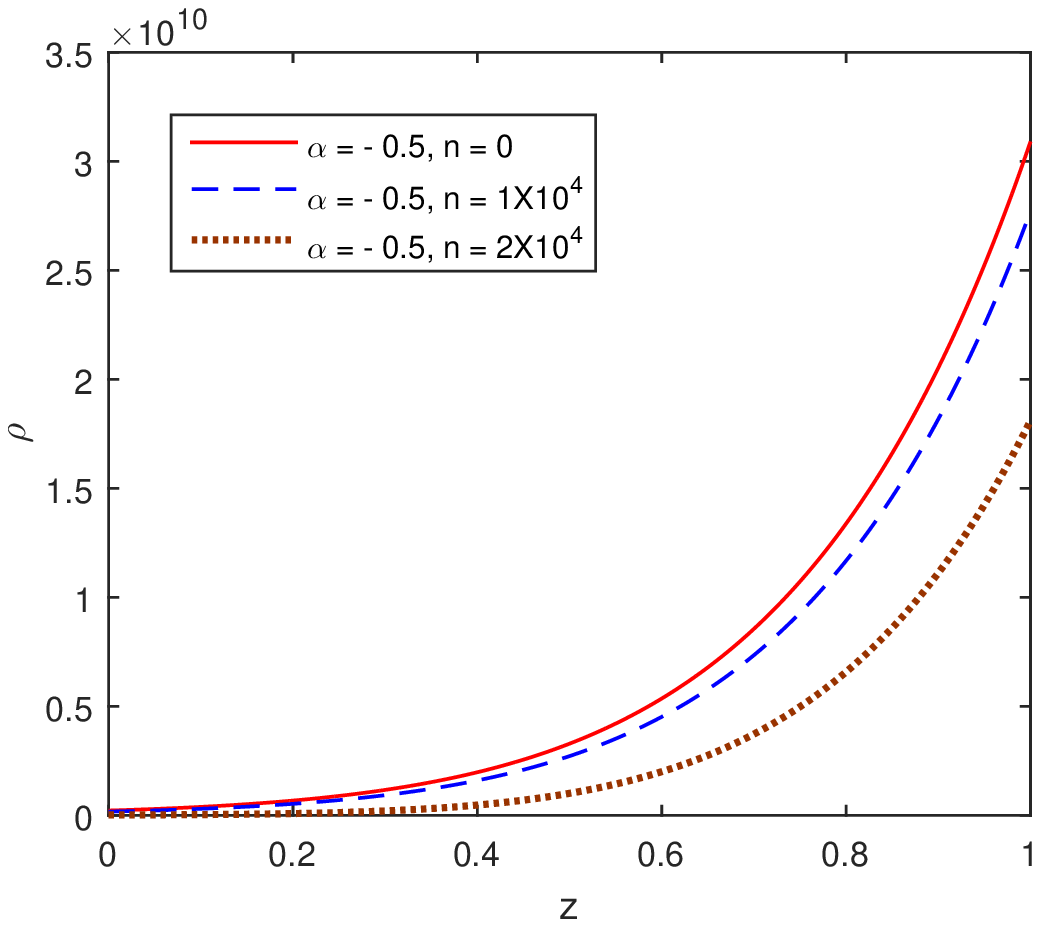}
	(c)\includegraphics[width=6cm,height=5cm,angle=0]{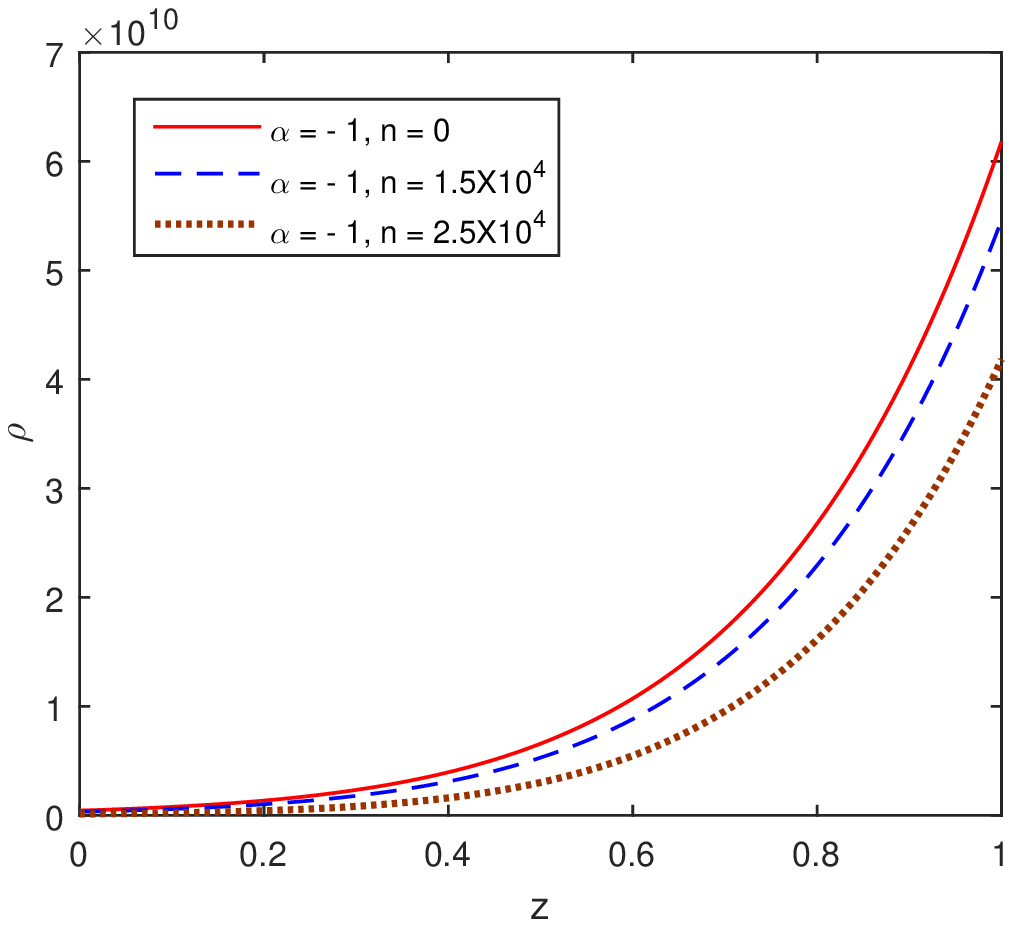}
	\caption{The plots of energy density $\rho(z)$ versus redshift $z$ for $H_{0}=71.27$.}
\end{figure}
\begin{figure}[H]
	(a)\includegraphics[width=5cm,height=5cm,angle=0]{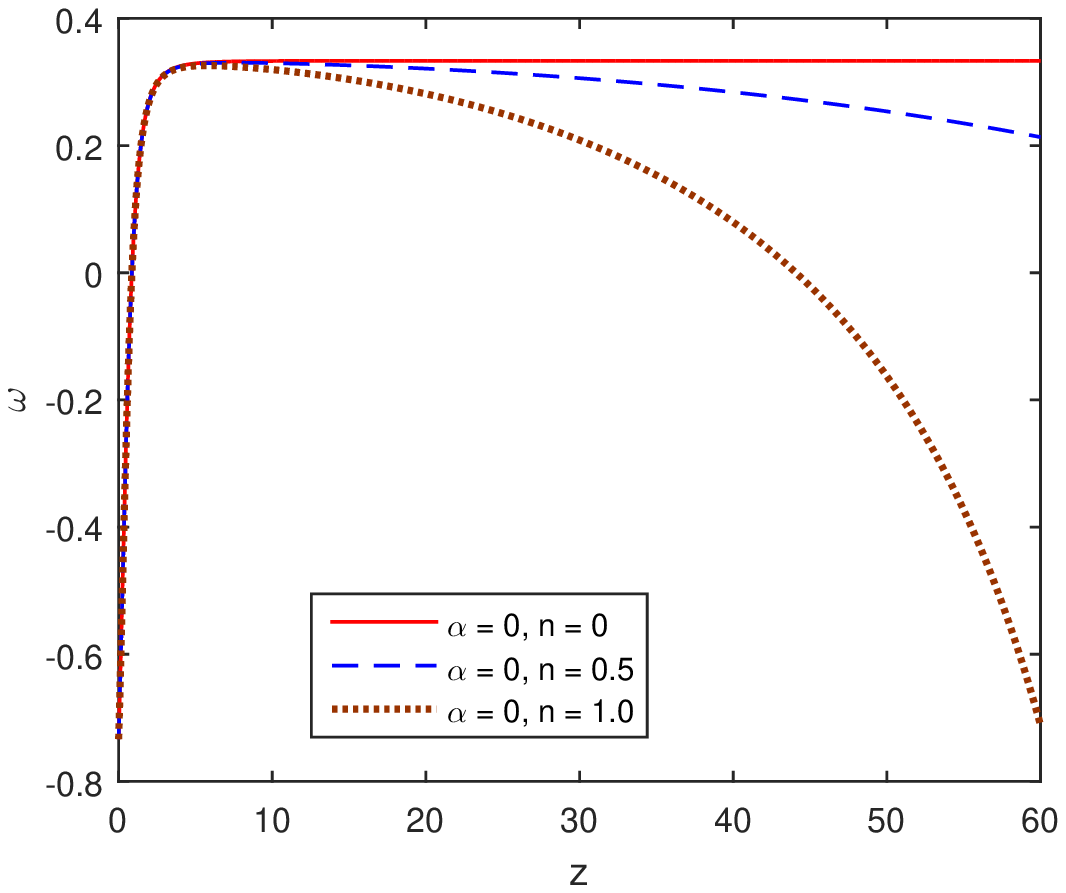}
	(b)\includegraphics[width=5cm,height=5cm,angle=0]{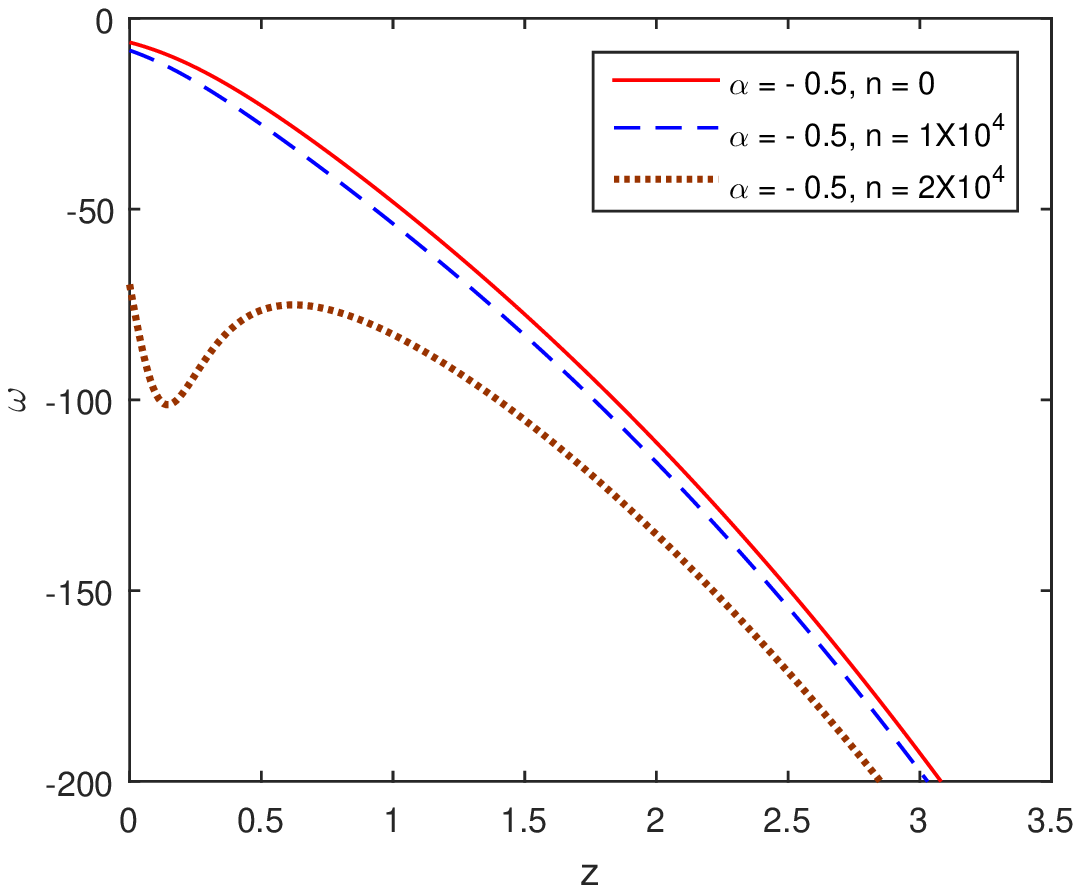}
	(c)\includegraphics[width=5cm,height=5cm,angle=0]{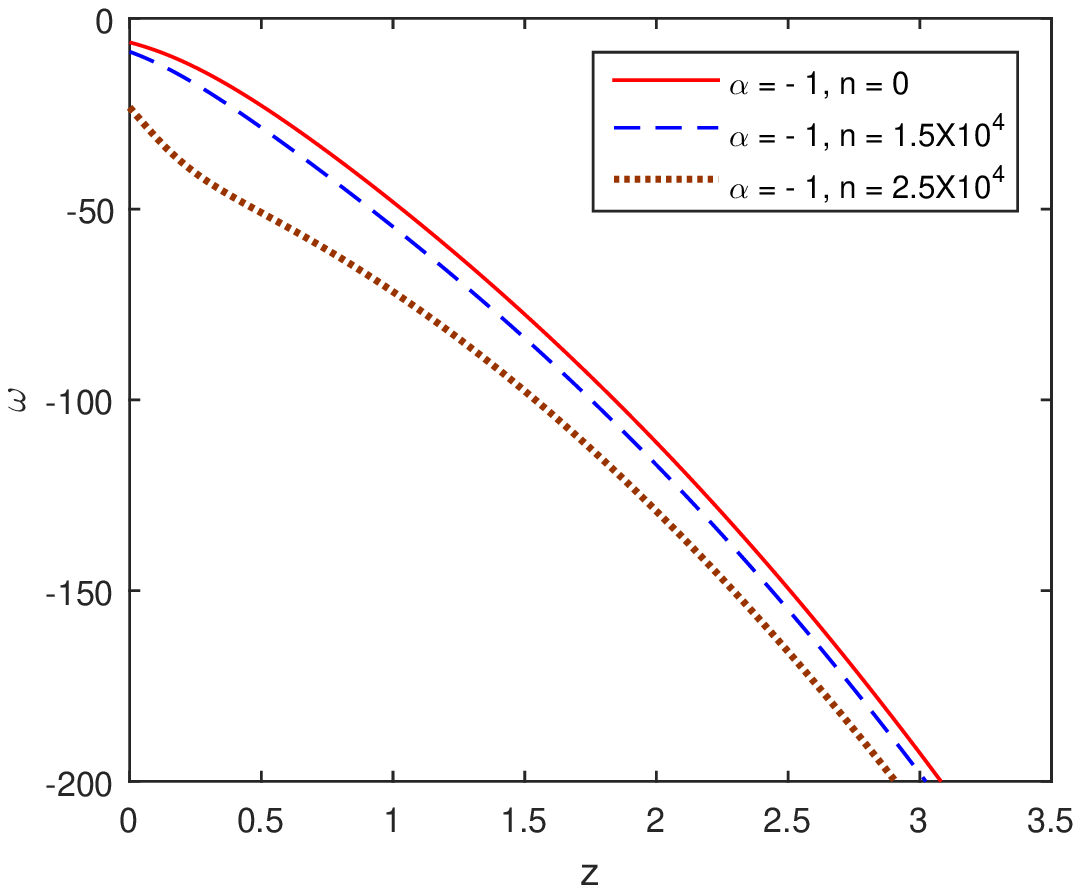}
	\caption{The plots of EoS parameter $\omega(z)$ versus redshift $z$ for $H_{0}=71.27$.}
\end{figure}
\begin{figure}[H]
	\centering
	\includegraphics[width=10cm,height=8cm,angle=0]{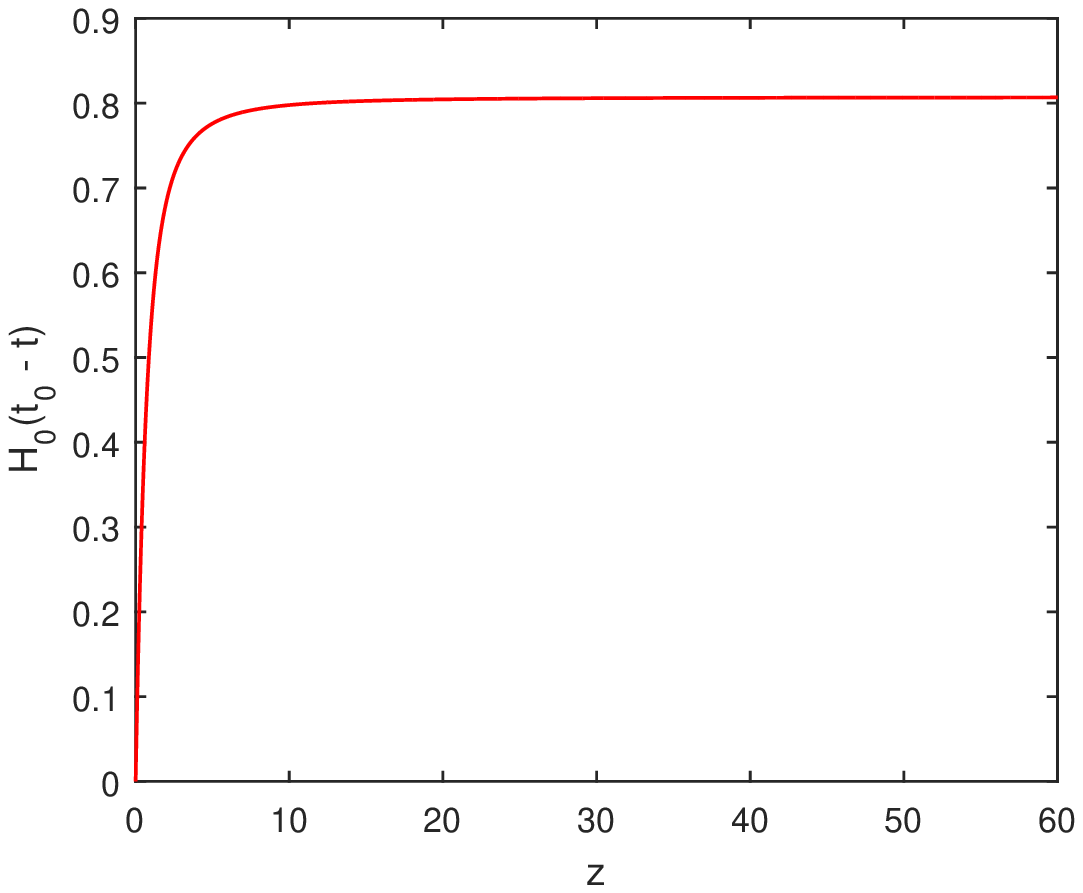}
	\caption{The plot of cosmic time $t$ versus redshift $z$}
\end{figure}
\subsection*{Age of the Universe}
The age of the cosmic universe is calculated as
\begin{equation}\label{46}
t_{0}-t=\int_{0}^{z}\frac{dz}{(1+z)H(z)}
\end{equation}
\begin{equation}\label{47}
(t_{0}-t)=\frac{1}{4}\sqrt{\frac{2}{c_{2}}}\left[ tanh^{-1}\left(\frac{4c_{2}-c_{1}}{4c_{2}}\right)^{\frac{1}{2}}-tanh^{-1}
\left(\frac{4c_{2}-c_{1}(1+z)^{4}}{4c_{2}}\right)^{\frac{1}{2}}  \right]  
\end{equation}
From Figure $11$, the constant graph of time shows the age of the universe i.e. $\lim\limits_{z\to\infty}H_{0}(t_{0}-t)=0.8233
\implies H_{0}t_{0}=0.8233$ and hence, the present age of the universe $t_{0}=11.3 Gyrs$.
\section{Concluding Remarks}
In this paper, we have investigated a flat FRW cosmological model filled with perfect fluid coupled with the zero-mass scalar field in the 
higher derivative theory of gravity. We have obtained two types universe models, first one is accelerating universe (power-law cosmology) 
and second one is transit phase model (hyperbolic expansion-law). The main features of the models are as:\\

\begin{itemize}
	\item The scale-factor $a(t)$ is obtained from the field equations without assumptions.
	\item These models evolves with high redshift values $0\leq z < \infty$.
	\item For the transit phase model the deceleration parameter shows signature-flipping with transition value $z_{t}=0.414$ 
	which is consistent with recent observations. The present value deceleration parameter is obtained as $q_{0}=-0.6$ for 
	transit model and $q_{0}=-0.5$ for accelerating universe.
	\item The EoS parameter $\omega$ varies from positive to negative values with the evolution the universe, which reveals the 
	formation of different structures during the evolution of the universe.
	\item At $t=0$, the scale factor $a(t)$ has a finite value.
	\item The present age of the universe for transit phase model is obtained as $11.3 Gyrs$.
\end{itemize}
\section*{Acknowledgment}
One of the authors (DM) is thankful to IASE (Deemed to be University), Sardarshahar, Rajsthan, India for providing facilities and support where part 
of this work is carried out.

\end{document}